\documentstyle[12pt,epsfig]{article}
\newcommand{\be}{\begin{equation}}
\newcommand{\en}{\end{equation}}
\newcommand{\beeq}{\begin{eqnarray}}
\newcommand{\eneq}{\end{eqnarray}}

\makeatletter
\def\vereq#1#2{\lower3pt\vbox{\baselineskip1.5pt \lineskip1.5pt
\ialign{$\m@th#1\hfill##\hfil$\crcr#2\crcr\sim\crcr}}}
\makeatother
\begin{document}

\begin{titlepage}
\begin{flushright}
{DPNU-02-39}
\end{flushright}

\vskip 1cm
\begin{center}
{\large\bf Gauge-Higgs unification in the 5 dimensional \\
 $E_6$ and $E_7$ GUTs on orbifold}

\vskip 1cm {\normalsize Naoyuki Haba$^{1,2}$ and Yasuhiro Shimizu$^2$}
\\
\vskip 0.5cm {\it $^1$Faculty of Engineering, Mie University, Tsu, Mie,
  514-8507, Japan}\\
\vskip 0.5cm {\it $^2$Department of Physics, Nagoya University, Nagoya, 
  464-8602, Japan}
\end{center}
\vskip .5cm

%%%%%%%%%%%%%%%%%%%%%%%%%%%%%%%%%%%%%%%%%%%%%%
%%%%%%%%%%%%%%  ABSTRACT %%%%%%%%%%%%%%%%%%%%%
%%%%%%%%%%%%%%%%%%%%%%%%%%%%%%%%%%%%%%%%%%%%%%
\begin{abstract}

We consider 5 dimensional $E_6$ and $E_7$ 
 supersymmetric 
 grand unified theories (GUTs) on an orbifold $S^1/Z_2$, 
 in which the Higgs fields arise from
 the 5th components of the 5D gauge fields.
We introduce matter fields in 5 dimensions, 
 then the Yukawa interactions are induced from 
 the gauge interactions 
 among the 5th component of gauge fields and 
 the bulk hyper-multiplets. 
The adjoint and fundamental 
 representations of bulk matter fields can induce 
 all Yukawa interactions 
 of the standard model in $E_6$ GUT. 
Furthermore, realistic fermion mass hierarchies and flavor
 mixings with the CP violating phase 
 can be reproduced by introducing the 
 4D brane localized fields and their interactions. 
$E_7$ GUT can produce all Yukawa interactions
 by introducing only adjoint matter hyper-multiplet
 in the bulk. 
The charge quantization and 
 anomaly free structure on the 4D wall 
 are satisfactory in our models.

\end{abstract}
\end{titlepage}
\setcounter{footnote}{0}
\setcounter{page}{1}
\setcounter{section}{0}
\setcounter{subsection}{0}
\setcounter{subsubsection}{0}

%%%%%%%%%%%%%%%%%%%%%%%%%%%%%%%%%%%%%%%%%%%%%%%%%%%%%%%%%%%%%%%%%%%
%%%%%%%%%%%%%%%%%%%%% INTRODUCTION %%%%%%%%%%%%%%%%%%%%%%%%%%%%%%%%
%%%%%%%%%%%%%%%%%%%%%%%%%%%%%%%%%%%%%%%%%%%%%%%%%%%%%%%%%%%%%%%%%%%
\section{Introduction}

Much attentions have been paid to gauge 
 theories in higher dimensions. 
Especially, grand unified theories (GUTs) in 
 higher dimensions on 
 orbifolds have been studied by many 
 authors\cite{5d,5d2,5d3,HMN}. 
One of the strongest motivations of 
 the higher dimensional gauge theory
 is the very 
 attractive idea that the gauge and the Higgs fields can be 
 unified in higher dimensions\cite{Manton:1979kb, Hosotani}.
Recently, this possibility has been revisited in 
 Ref.\cite{Krasnikov:dt,Hall:2001zb,Burdman:2002se}.
In these scenarios the Higgs fields, which give masses 
 for quarks and leptons, are identified with the
 components of the gauge fields in higher dimensions.
The masses of Higgs fields are forbidden by 
 the higher dimensional gauge invariance, 
 and the supersymmetric non-renormalization theorem 
 protects Higgs fields from getting radiative
 induced masses. 
This is the reason why this scenario has
 light Higgs fields where
 the gauge group in higher dimensions must be lager than the 
 standard model (SM) gauge
 group in order to obtain the Higgs doublets from the gauge fields
 in higher dimensions. 
The gauge symmetries can be reduced 
 by the orbifolding boundary conditions of extra 
 dimensions.

For the identification of Higgs fields as 
 a part of the gauge super-multiplet, 
 there are two ways. 
One is considering 6D $N=2$ supersymmetric 
 gauge theory, 
 whose 5th and 6th coordinates are compactified
 on $T^2/(Z_2 \times Z_2')$ orbifold\cite{Hall:2001zb}. 
This theory corresponds to 4D $N=4$ supersymmetric gauge theory,
 where there are chiral superfields in the gauge
 super-multiplet 
 which do not contain components of the 
 higher-dimensional gauge bosons, and thus are 
 able to couple with 4D localized matter. 
Higgs fields are identified as the part of 
 these chiral superfields, 
 which can have gauge invariant 
 Yukawa interactions in the 4D superpotential.
Another is considering 5D $N=1$ supersymmetric 
 gauge theory 
 whose 5th coordinate is compactified
 on $S^1/Z_2$ orbifold\cite{Burdman:2002se}.  
This theory corresponds to 4D $N=2$ supersymmetric gauge theory,
 where chiral superfields 
 in the gauge super-multiplet transform 
 nonlinearly under the gauge transformation, 
 and thus can not couple with 
 the 4D brane localized matters. 
However, if we 
 introduce the matter fields in the bulk, 
 and identify 
 Higgs fields as the 5D components of 
 the gauge super-multiplet, 
 there can appear ``Yukawa'' interactions 
 among the bulk matters and ``Higgs'' fields.  
In Ref.\cite{Burdman:2002se} they have
 considered the 
 Higgs doublets are induced 
 as the zero modes of the 5th components of the 
 gauge super-multiplet. 
However, 
 in order to obtain all Yukawa interactions 
 of quarks and leptons, 
 they must introduce a lot of matter 
 representations in the bulk. 
Moreover,  
 the charge quantization is 
 not automatic since 
 the fixed 4D wall does not have 
 GUT gauge symmetry, and 
 the anomaly free structure is 
 not simple.

In this paper, we 
 consider the 5D 
 $E_6$ and $E_7$ 
 supersymmetric GUTs on 
 an orbifold, $S^1/Z_2$. 
We adopt such large gauge groups for 
 respecting  
 the charge quantization and 
 the simple anomaly free content on 4D fixed walls. 
A small numbers of matter representations 
 are enough for inducing all Yukawa 
 interactions of the SM. 
By imposing the non-trivial parity and boundary conditions for
 the gauge fields, the gauge groups are reduced to their subgroups.
Furthermore, we show that the 
 suitable Higgs fields can arise from the 
 5th components of the bulk gauge fields. 
We introduce matter fields in 5 dimensions 
 so that the Yukawa interactions are induced from 
 the gauge interactions 
 among the 5th component of gauge fields, 
 which will be identified as the Higgs fields, and 
 the hyper-multiplets in the bulk. 
In $E_6$ GUT, 
 we will introduce the adjoint and fundamental 
 matter representations in 
 order to generate all Yukawa interactions 
 of quarks and leptons. 
We will introduce the additional 
 matter and Higgs fields on the 4D brane 
 with anomaly free contents. 
The vacuum expectation values (VEVs) of the 
 brane localized Higgs fields 
 make unwanted matter fields heavy as well as 
 generate effective 4D Yukawa interactions. 
For the realistic fermion 
 mass hierarchies and flavor mixings\cite{flavor,HKS2}, 
 our scenario can induce not only 
 the suitable mass hierarchies and flavor mixings  
 but also the CP violating phase 
 through the brane 
 localized interactions. 
In case of 
 $E_7$ GUT, 
 all Yukawa interactions can be produced 
 by only adjoint representation matter 
 hyper-multiplets in the bulk.

%%%%%%%%%%%%%%%%%%%%%%%%%%%%%%%%%%%%%%%%%%%%%%%%%%%%%%%%%%%%%%%%%%%
%%%%%%%%%%%%%            Section            %%%%%%%%%%%%%%%%%%%%%%%
%%%%%%%%%%%%%%%%%%%%%%%%%%%%%%%%%%%%%%%%%%%%%%%%%%%%%%%%%%%%%%%%%%%
\section{Gauge-Higgs sector in 5D model}

At first let us show the notation of 
 the 5D $N=1$ SUSY theory, 
 which corresponds to 4D $N=2$ SUSY one\cite{6DLag}. 
The vector multiplet, $(V,\Sigma)$, of the 
 $N=2$ SUSY gauge theory 
 is written as 
\begin{eqnarray}
&&V=-\theta \sigma^m \bar{\theta}A_m + i \bar{\theta}^2 \theta \lambda
    -i \theta^2  \bar{\theta} \bar{\lambda} +
    {1\over2}\bar{\theta}^2\theta^2D ,\\
&&\Sigma={1\over\sqrt{2}}(\sigma+iA_5)+\sqrt{2}\theta \lambda' + 
 \theta^2F.
\end{eqnarray}
In the non-abelian gauge theory,
 the gauge transformation 
 is given by 
$e^V \rightarrow  h^{-1}V\bar{h}^{-1}$ and  
$\Sigma \rightarrow h^{-1}(\Sigma + \sqrt{2}\partial_y) h$,
where we denote 
$V\equiv V^aT^a$, $\Sigma \equiv \Sigma^aT^a$ and 
$h\equiv e^{-\Lambda}$, $\bar{h}\equiv e^{-\bar{\Lambda}}$. 
Then the action is given by
\begin{eqnarray}
S_{5D}&=&\int d^4x dy \left[{1\over4kg^2}{\rm Tr}\left\{
  \int d^2 \theta W^\alpha W_\alpha
     +h.c.\right\} \right. \nonumber \\ 
&& + \left. \int d^2 \theta {1\over kg^2}{\rm Tr}\left( 
     (\sqrt{2}\bar{\partial}_y  + \bar{\Sigma})e^{-V}
     (-\sqrt{2}\partial_y +{\Sigma})e^V
     +\bar{\partial}e^{-V}\partial e^V
     \right)     \right],
\end{eqnarray}
where ${\rm Tr}(T^aT^b)=k\delta_{ab}$. 
We denote $y$ as the 5th dimensional coordinate.

As for the hyper-multiplet, $(H,H^c)$, 
 they transform $H \rightarrow hH$ and $H^c \rightarrow h^c{}^{-1}H^c$ 
 under the gauge transformation, 
 where $h=e^{-\Lambda^a T^a}$ and
 $h^c=(h^{-1})^T=(e^{\Lambda^a T^a})^T$. 
The action of them is given by
\begin{eqnarray}
S_{5D}^H &=& \int d^5x dy \left[
  \int d^4 \theta (H^c e^V \bar{H}^c + \bar{H}e^{-V}H)+ \right. \nonumber \\ 
&& + \left. \left[\int d^2 \theta 
   \left( H^c \left(m+\left(\partial_y-{1\over\sqrt{2}}\Sigma \right)
  \right)H\right)  +h.c. \right]\right].
\label{5D}
\end{eqnarray}
This means that $\Sigma$ must change the sign 
 under the $Z_2$ projection, $P: y \rightarrow -y$, 
 and bulk constant mass $m$ is forbidden\footnote{
 When the $y$ dependent mass $m(-y)=-m(y)$ is introduced, 
 it makes the zero mode wave-function in the bulk 
 be localized at $y=0$\cite{Kaplan:2001ga}.}.
The field $H^c$ is so-called mirror field, 
 which should be the odd eigenstate of $P$, since 
 it is the right-handed field. 
Thus, 
 the parity operator $P$ acts on fields as
\beeq
&& V(y)=P V(-y) P^{-1}, \;\;\;\;\;\; \Sigma(y)= - P \Sigma(-y) P^{-1}, 
\label{5}\\
&& H(y)=P H(-y), \;\;\;\;\;\;\;\;\;\;\; H^c(y)= - P H^c(-y). 
\label{6}
\eneq
As for the boundary condition, $T$, 
 the bulk 
 fields are transform as 
\beeq
&& V(y)=T V(y+2\pi R) T^{-1}, \;\;\;\;\;\; 
   \Sigma(y)= T \Sigma(y+2\pi R) T^{-1}, 
\label{8} \\
&& H(y)=T H(y+2\pi R), \;\;\;\;\;\;\;\;\;\;\; 
   H^c(y)= T H^c(y+2\pi R). 
\label{9}
\eneq
We will consider the cases where 
 $P$ and $T$ have the nontrivial eigenvalues 
 in the GUT bases.

We can show that 
 the $F$-term interaction in Eq.(\ref{5D}) 
\begin{eqnarray}
W_Y \supset H^c \Sigma H,
\label{Yukawa}
\end{eqnarray}
is permitted under the $Z_2$ projection. 
This interaction connects the chiral and mirror
 fields through the chirality flip, which
 seems to be just the Yukawa interaction 
 in the 4D theory. 
This becomes justified 
 when $\Sigma$ has zero modes at the components 
 of the fundamental Higgs fields.  
We will find 
 $E_6$ and $E_7$ GUTs where 
 Yukawa interactions 
 are really coming from Eq.(\ref{Yukawa}).

%%%%%%%%%%%%%%%%%%%%%%%%%%%%%%%%%%%%%%%%%%%%%%%%%%%%%%%%%%%%%%%%%%%
%%%%%%%%%%%%%            Section            %%%%%%%%%%%%%%%%%%%%%%%
%%%%%%%%%%%%%%%%%%%%%%%%%%%%%%%%%%%%%%%%%%%%%%%%%%%%%%%%%%%%%%%%%%%
\section{$E_6$ GUT}

At first let us consider the $E_6$ GUT. 
The adjoint representation of {\bf 78} of $E_6$ is 
 divided as,
\beeq
&&{\bf 78 = 45_0 + 1_0 + 16_{-3}+\overline{16}_3},
                          \;\;\;\;\;\;\;\;SO(10)\times U(1)_X, 
\label{10} \\
&&{\bf 78 = (35,1)+(1,3)+(20,2)},
                          \;\;\;\;\;\;SU(6)\times SU(2). 
\label{11} 
\eneq
We take the parity $P$ in Eqs.(\ref{5}) and (\ref{6}) 
 and the boundary condition $T$ in Eqs.(\ref{8}) and (\ref{9}).

As for vector hyper-multiplets, 
 we take $P=+1$ for ${\bf 45_0 + 1_0}$, 
 and $P=-1$ for 
 ${\bf 16_{-3}+\overline{16}_3}$ in Eq.(\ref{10}). 
On the other hand, 
 we take $T=+1$ for ${\bf (35,1)+(1,3)}$, and 
 $T=-1$ for 
 ${\bf (1,3)+(20,2)}$ in Eq.(\ref{11}). 
These $P$ and $T$ make 
 the $E_6$ gauge symmetry reduce to 
 $SU(5) \times U(1)_V \times U(1)_X$. 
It is because 
 the vector superfield, $V_{\bf 78}$, is divided into 
\beeq
&&V_{\bf 78}{\bf =24_{(0,0)}^{(+,+)}+1_{v(0,0)}^{(+,+)}+1_{x(0,0)}^{(+,+)}
       +10_{(4,0)}^{(+,-)}+\overline{10}_{(-4,0)}^{(+,-)}
       +10_{(-1,-3)}^{(-,-)}+\overline{10}_{(1,3)}^{(-,-)}} \nonumber \\
&&\;\;\;\;\;\;\;
      {\bf +\overline{5}_{(3,-3)}^{(-,+)}+5_{(-3,3)}^{(-,+)}
       +1_{(-5,-3)}^{(-,+)}+1_{(5,3)}^{(-,+)}},
\label{78}
\eneq
under $SU(5) \times U(1)_V \times U(1)_X$. 
where $(\pm, \pm)$ represents the eigenvalues of $(P,T)$. 
The gauge reduction, 
 $E_6 \rightarrow SU(5) \times U(1)_V \times U(1)_X$, 
 can be easily seen by the zero mode elements  
 $(+,+)$ of $V$. 
On the other hand, 
 the chiral superfield 
 $\Sigma_{\bf 78}$ is divided into 
\beeq
&&\Sigma_{\bf 78}{\bf =24_{(0,0)}^{(-,+)}+1_{v(0,0)}^{(-,+)}+1_{x(0,0)}^{(-,+)}
       +10_{(4,0)}^{(-,-)}+\overline{10}_{(-4,0)}^{(-,-)}
       +10_{(-1,-3)}^{(+,-)}+\overline{10}_{(1,3)}^{(+,-)}} \nonumber \\
&&\;\;\;\;\;\;\;
      {\bf +\overline{5}_{(3,-3)}^{(+,+)}+5_{(-3,3)}^{(+,+)}
       +1_{(-5,-3)}^{(+,+)}+1_{(5,3)}^{(+,+)}},
\label{78s}
\eneq
under $SU(5) \times U(1)_V \times U(1)_X$. 
The zero modes, $(+,+)$, suggest that
 there appear the following four Higgs fields, 
\be
{\bf \overline{5}_{(3,-3)}^{H}, \; 5_{(-3,3)}^{H}, \;
       1_{(-5,-3)}^{H}, \; 1_{(5,3)}^{H}},
\label{14}
\en
in the low energy, 
 since we regard $\Sigma$ as the Higgs fields
 as in Eq.(\ref{Yukawa}).

{}For realizing the suitable gauge reduction,
 $E_6 \rightarrow SU(5) \times U(1)_V \times U(1)_X$, 
 there are two other choices of 
 $P$ and $T$. 
One is taking $P=+1$ for 
 ${\bf (35,1)+(1,3)}$ and 
 $P=-1$ for ${\bf (1,3)+(20,2)}$ in Eq.(\ref{11}), and 
 $T=+1$ for ${\bf 45_0 + 1_0}$ and $T=-1$ for 
 ${\bf 16_{-3}+\overline{16}_3}$ in Eq.(\ref{10}). 
$V$ has zero modes in the components of 
 ${\bf 24_{(0,0)}+1_{v(0,0)}+1_{x(0,0)}}$, while 
 $\Sigma$ has zero modes at 
 ${\bf 10_{(4,0)}+\overline{10}_{(-4,0)}}$ 
 in Eq.(\ref{78s}). 
The other is taking $P=+1$ for 
 ${\bf 45_0 + 1_0}$ and 
 $P=-1$ for ${\bf 16_{-3}+\overline{16}_3}$, 
 and $P'=+1$ for ${\bf (35,1)+(1,3)}$ and
 $P'=-1$ for ${\bf (20,2)}$, 
 where $P' \equiv TP$
 is the 
 reflection around $y=\pi R$\cite{5d3}.
This case suggests 
 the residual gauge symmetry at 
 $y=0$ is $SO(10)\times U(1)_X$ 
 and that at $y=\pi R$ is $SU(6)\times SU(2)$.  
This case has zero mode components at 
 ${\bf 24_{(0,0)}+1_{v(0,0)}+1_{x(0,0)}}$ in $V$, 
 while $\Sigma$ has zero modes at 
 ${\bf 10_{(-1,-3)}+\overline{10}_{(1,3)}}$ 
 in Eq.(\ref{78s}). 
Since ${\bf 10 + \overline{10}}$ representations of 
 $SU(5)$ can not 
 be the Higgs fields of the SUSY SM, 
 these two cases are not suitable for the 
 phenomenology. 
Therefore, the choice of $P$ and $T$ in 
 Eqs.(\ref{78}) and (\ref{78s})  
 is the unique one for obtaining 
 the suitable Higgs fields. 
{}From now on we take $P$ and $T$ as 
 Eqs.(\ref{78}) and (\ref{78s}) in order to 
 regard zero modes fields in Eq.(\ref{14}) 
 as the Higgs fields.

\begin{figure}
\begin{center}
\epsfig{file=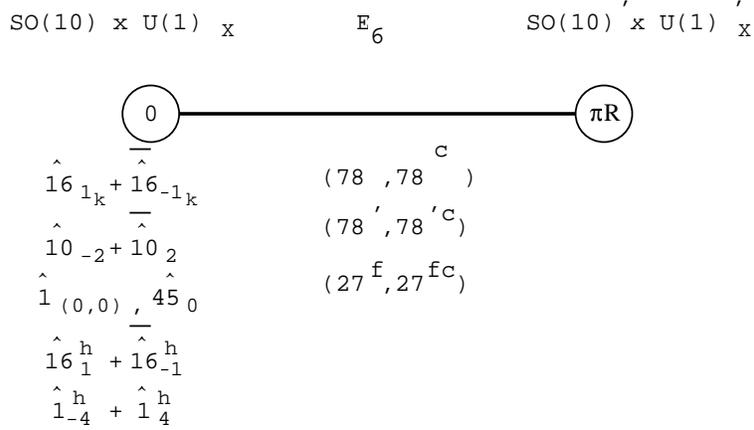,width=10cm}
\caption{Matter configuration on the $S^1/Z_2$ orbifold in the $E_6$
 GUT. }
\label{fig:E6}
\end{center}
\end{figure}

Next let us discuss the matter field contents. 
In Fig.\ref{fig:E6} we show the matter configuration
 on the orbifold.
We introduce three bulk hyper-multiplets, 
 which are two adjoint representations, 
 ${\bf (78,78^c)}$ and ${\bf (78',78'^c)}$, 
 and one fundamental representation, 
 ${\bf (27,27^c)}$. 
The bulk has the $E_6$ gauge symmetry. 
There are two fixed 4D walls at $y=0$ and $y=\pi R$, 
 where 
 gauge symmetries are $SO(10)\times U(1)_X$
 and $SO(10)'\times U(1)'_X$, respectively.
$SO(10)' \times U(1)'_X$ is contained in $E_6$ 
 in the different base from $SO(10)\times U(1)_X$. 
Since we can input 
 the fields with the representations of 
 reduced gauge symmetry 
 on the 4D walls\cite{HMN},  
 we introduce chiral superfields,
 $\hat{\bf 16}_{\bf -3} + \overline{\hat{\bf 16}}_{\bf -3}$, 
 $2\times (\hat{\bf 16}_{\bf 1} + \overline{\hat{\bf 16}}_{\bf -1})$, 
 and $\hat{\bf 45_0}$ at $y=0$,
 where we are living. 
We will also introduce vector-like 
 4D localized Higgs fields, 
 $\hat{\bf 16}^h_{\bf 1} + \overline{\hat{\bf 16}}^h_{\bf -1}$ and 
 $\hat{\bf 1}^h_{\bf -4} + \hat{\bf 1}^h_{\bf 4}$ at $y=0$. 
We will see the details of the matter contents below.

As for the bulk matter fields, 
 matter hyper-multiplet, $(H,H^c)$, can have 
 all combinations of $(P,T)=(\pm, \pm)$ 
 contrary to the gauge super-multiplets. 
But, we should notice that 
 once the eigenvalues of $(P,T)$ of $H$ are given, 
 those of $H^c$ are automatically determined. 
As for the adjoint representation matter 
 hyper-multiplets, 
 we introduce ${\bf (78,78^c)}$ and ${\bf (78',78'^c)}$, 
 which have all combinations of eigenvalues of $(P,T)$ 
 in their components. 
The ${\bf (78,78^c)}$ is divided as 
\beeq
&&{\bf 78 =24_{(0,0)}^{(-,-)}+1_{v(0,0)}^{(-,-)}+1_{x(0,0)}^{(-,-)}
       +10_{(4,0)}^{(-,+)}+\overline{10}_{(-4,0)}^{(-,+)}
       +10_{(-1,-3)}^{(+,+)}+\overline{10}_{(1,3)}^{(+,+)}} \nonumber \\
&&\;\;\;\;\;\;\;
      {\bf +\overline{5}_{(3,-3)}^{(+,-)}+5_{(-3,3)}^{(+,-)}
       +1_{(-5,-3)}^{(+,-)}+1_{(5,3)}^{(+,-)}}, 
\label{24} \\
&&{\bf 78^c =24_{(0,0)}^{(+,-)}+1_{v(0,0)}^{(+,-)}+1_{x(0,0)}^{(+,-)}
       +10_{(4,0)}^{(+,+)}+\overline{10}_{(-4,0)}^{(+,+)}
       +10_{(-1,-3)}^{(-,+)}+\overline{10}_{(1,3)}^{(-,+)}} \nonumber \\
&&\;\;\;\;\;\;\;
      {\bf +\overline{5}_{(3,-3)}^{(-,-)}+5_{(-3,3)}^{(-,-)}
       +1_{(-5,-3)}^{(-,-)}+1_{(5,3)}^{(-,-)}}, 
%\label{25}
\eneq
and ${\bf (78',78'^c)}$ is divided as 
\beeq
&&{\bf 78' ={24'}_{(0,0)}^{(-,+)}+{1'}_{v(0,0)}^{(-,+)}+{1'}_{x(0,0)}^{(-,+)}
       +{10'}_{(4,0)}^{(-,-)}+\overline{10}'{}_{(-4,0)}^{(-,-)}
       +{10'}_{(-1,-3)}^{(+,-)}+\overline{10}'{}_{(1,3)}^{(+,-)}} \nonumber \\
&&\;\;\;\;\;\;\;
      {\bf +\overline{5}'{}_{(3,-3)}^{(+,+)}+5'{}_{(-3,3)}^{(+,+)}
       +1'{}_{(-5,-3)}^{(+,+)}+1'{}_{(5,3)}^{(+,+)}}, 
\label{26} \\
&&{\bf 78'{}^c =24'{}_{(0,0)}^{(+,+)}+1'{}_{v(0,0)}^{(+,+)}
                                       +1'{}_{x(0,0)}^{(+,+)}
       +10'{}_{(4,0)}^{(+,-)}+\overline{10}'{}_{(-4,0)}^{(+,-)}
       +10'{}_{(-1,-3)}^{(-,-)}+\overline{10}'{}_{(1,3)}^{(-,-)}} \nonumber \\
&&\;\;\;\;\;\;\;
      {\bf +\overline{5}'{}_{(3,-3)}^{(-,+)}+5'{}_{(-3,3)}^{(-,+)}
       +1'{}_{(-5,-3)}^{(-,+)}+1'{}_{(5,3)}^{(-,+)}}, 
\label{27}
\eneq
under $SU(5) \times U(1)_V \times U(1)_X$. 
Therefore, from Eqs.(\ref{78s}), (\ref{24}), (16), 
 (\ref{26}), and (\ref{27}), the Yukawa interaction in 
 Eq.(\ref{Yukawa}) becomes 
\beeq
&&W_Y^a \supset{\bf 78^c}\; \Sigma_{\bf 78}\; {\bf 78}\; +
  {\bf 78'{}^c}\; \Sigma_{\bf 78'}\; {\bf 78'}  \nonumber \\
&&\simeq{\bf 10_{(4,0)}\; 5_{(-3,3)}^H \; 10_{(-1,-3)}+
  \overline{10}_{(-4,0)}\; \overline{5}_{(3,-3)}^H \; 
                               \overline{10}_{(1,3)}}\nonumber \\
&&{\bf +[24'{}_{(0,0)}+1'{}_{v(0,0)}+1'{}_{x(0,0)}] 
           5_{(-3,3)}^H  \overline{5}'{}_{(3,-3)} 
       +[24'{}_{(0,0)}+1'{}_{v(0,0)}+1'{}_{x(0,0)}] 
           \overline{5}_{(3,-3)}^H  5'{}_{(-3,3)}}\nonumber \\ 
&&{\bf +[1'{}_{v(0,0)}+1'_{x(0,0)}] 1_{(5,3)}^H \; 1'{}_{(-5,-3)}
       +[1'{}_{v(0,0)}+1'_{x(0,0)}] 1_{(-5,-3)}^H \; 1'{}_{(5,3)}} ,
\label{y21}
\eneq
by writing only the zero modes.  
On the other hand, 
 we introduce 
 the fundamental representation 
 of matter hyper-multiplet, ${\bf (27^f, 27^f{}^c)}$,  
 for the Yukawa interactions of the 
 down-sector quarks and the charged leptons\footnote{
 The higher order operator, 
 ${1\over M_{*}}[{\bf 78'}^\dagger  {\bf 78}]_D$, 
 can not induce the Yukawa interactions of the 
 down-sector quarks and the charged leptons, 
 where $M_{*}$ is the cut-off scale. 
It is because 
 the Higgs fields are originated in 
 ${\bf 16^H+\overline{16}^H}$ representations, 
 and matter fields (${\bf 10}$ and ${\bf \overline{5}}$ in $SU(5)$)
 are originated in {\bf 45} or {\bf 16} under 
 $SO(10)$ in this model. 
Thus, there are no Yukawa interactions 
 of down-quark and charged lepton sectors 
 in this higher order operator, since  
 we can not make singlet by ${\bf 16\;16^H\;16}$ nor 
 ${\bf 45\;16^H\;16}$. 
}. 
The fundamental representation is 
 divided into
\beeq
&&{\bf 27^f = 16^f_1 + 10^f_{-2} + 1^f_4}, \;\;\;\; 
  {\bf 27^f{}^c = \overline{16}_{-1}^f + \overline{10}_2^f + 1_{-4}^f}, 
                          \;\;\;SO(10)\times U(1)_X, \\
&&{\bf 27^f = (15^f,1^f)+(\overline{6}^f,2^f)}, \;
  {\bf 27^f{}^c = (\overline{15}^f,1^f)+(6^f,2^f)}, 
                          \;\;\;SU(6)\times SU(2). 
\eneq
By choosing the eigenvalues of $P$ and $T$, 
 ${\bf (27,27^c)}$ is divided into 
\beeq
&&{\bf 27^f = 10_{(-1,1)}^{f(+,+)} + \overline{5}_{(3,1)}^{f(+,-)} 
           + 1_{(-5,1)}^{f(+,-)}+5_{(2,-2)}^{f(-,+)}
           + \overline{5}_{(-2,-2)}^{f(-,-)}+1_{(0,4)}^{f(-,-)}}, 
\label{19}\\
&&{\bf 27^f{}^c = \overline{10}_{(1,-1)}^{f(-,+)} + 5_{(-3,-1)}^{f(-,-)} 
           + 1_{(5,-1)}^{f(-,-)}+\overline{5}_{(-2,2)}^{f(+,+)}
           + 5_{(2,2)}^{f(+,-)}+1_{(0,-4)}^{f(+,-)}}, 
\label{20}
\eneq
under $SU(5)\times U(1)_V \times U(1)_X$. 
This suggests that 
 the Yukawa interaction Eq.(\ref{Yukawa}) becomes 
\beeq
W_Y^f \supset {\bf 27^f{}^c}\; \Sigma_{\bf 78}\; {\bf 27^f} \simeq 
  {\bf \overline{5}_{(-2,2)}^f \; \overline{5}_{(3,-3)}^{H}\; 
                                                  10_{(-1,1)}^f},
\label{27s}
\eneq
in the low energy.

{}For 
 the 
 brane-localized 
 matter fields, 
 $2 \times (\hat{\bf 16}_{\bf 1}+\overline{\hat{\bf 16}}_{\bf -1})$, 
 $\hat{\bf 10}_{\bf -2}+ {\hat{\bf 10}}_{\bf -2}$, 
 $\hat{\bf 1}_{\bf (0,0)}$, 
 and 
 $\hat{\bf 45_0}$, 
 and 
 brane-localized Higgs fields, 
 $\hat{\bf 16}^h_{\bf 1} + \overline{\hat{\bf 16}}^h_{\bf -1}$ and 
 $\hat{\bf 1}^h_{\bf -4} + \hat{1}^h_{\bf 4}$,
 there is the 
 superpotential $W_{4D}^m$ at 
 $y=0$ 4D wall, which is given by 
\beeq
W_{4D}^m&=& 
     \sum_{k=1,2}\overline{\hat{\bf 16}}_{{\bf -1}_k}
     (m_{{10}_k} \hat{\bf 1}_{\bf 4}^h \; {\bf 16_{-3}}+ 
      m_{{10^c}_k} \hat{\bf 16}^h_{\bf 1} {\bf 45_0} + 
      m_{{10^f}_k}{\bf 16_1^f})  \nonumber \\
&+&  \sum_{k=1,2}{\hat{\bf 16}}_{{\bf 1}_k}
     (\bar{m}_{{10}_k} \hat{\bf 1}_{\bf -4}^h \; 
              \overline{\bf 16}_{\bf 3}+ 
      \bar{m}_{{10^c}_k} \overline{\hat{\bf 16}}^h_{\bf -1} 
             {\bf 45_0}) \nonumber \\ % + 
&+& \overline{\hat{\bf 10}}_{\bf 2}
     (m_{5'} \hat{\bf 16}^h_{\bf 1}{\bf 16_{-3}}+ 
      m_{5^f}{\bf  10_{-2}^f})+  
     \hat{\bf 1}_{\bf 0} 
   (m_{1'{}^c} \hat{\bf 1}_{\bf 4}^h \; \overline{\hat{\bf 16}}^h_{\bf -1}
       \; {\bf 16_{-3}'}
  + m_{1'} {\bf 1'{}_{0}}) \nonumber \\
&+& \bar{m}_{5'} 
  {\hat{\bf 10}}_{\bf -2} \overline{\hat{\bf 16}}^h_{\bf -1}
                 \overline{\bf 16}_{\bf 3}+ 
    \bar{m}_{1'{}^c} 
    \hat{\bf 1}_{\bf 0} 
    {\hat{\bf 1}}_{\bf -4}^h \; 
                       {\hat{\bf 16}}^h_{\bf 1}
       \; \overline{\bf 16}_{\bf 3}' 
    +m_{24}\; \hat{\bf 45_0}\;{\bf 45_0}  
\label{25} \\
&+&    \sum_{k=1,2}M_{{10}_k} \overline{\hat{\bf 16}}_{{\bf -1}_k} 
             {\hat{\bf 16}}_{{\bf 1}_k} 
      + M_5 {\hat{\bf 10}}_{\bf 2}{\hat{\bf 10}}_{\bf -2}
      + M_1 \hat{\bf 1}_{\bf 0}^2       
      + M_{24}\; \hat{\bf 45_0}^2 ,
\nonumber 
\eneq
where $k=1,2$. 
In Eq.(\ref{25}), 
 the mass dimension $M_{*}$ is suppressed. 
We assume that the mass term 
 $m_{24}(\overline{\hat{\bf 10}}_{\bf (-4,0)}{\bf 10_{(4,0)}}
 +\hat{\bf 10}_{\bf (4,0)}\overline{\bf 10}_{\bf (-4,0)})$, 
 which is a part of $m_{24}\hat{\bf 45_0}{\bf 45_0}$, 
 is forbidden by the underlying theory,  
 while 
 the mass term 
 $m_{24}({\hat{\bf 24}}_{\bf (0,0)}{\bf 24_{(4,0)}}
 +\hat{\bf 1}_{\bf v(0,0)}{\bf 1_{v(0,0)}})$ 
 exists at $y=0$ wall.

When the 4D Higgs fields, 
 $\hat{\bf 16}^h_{\bf 1} + \overline{\hat{\bf 16}}^h_{\bf -1}$ and 
 $\hat{\bf 1}^h_{\bf -4} + \hat{\bf 1}^h_{\bf 4}$, 
 take 
 VEVs,
\beeq
 \langle \hat{\bf 16}^h_{\bf 1} \rangle = 
               \langle \hat{\bf 1}^h_{\bf (-5,1)} \rangle , \;\;
   \langle \overline{\hat{\bf 16}}^h_{\bf -1} \rangle =
                     \langle \hat{\bf 1}^h_{\bf (5,-1)} \rangle ,  \;\;
 \langle \hat{\bf 1}^h_{\bf -4} \rangle = 
                    \langle \hat{\bf 1}^h_{\bf (0,-4)} \rangle , \;\;
   \langle \hat{\bf 1}^h_{\bf 4} \rangle =
                     \langle \hat{\bf 1}^h_{\bf (0,4)} \rangle ,
\label{32}
\eneq
in the bases of $SU(5) \times U(1)_V \times U(1)_X$, 
 the gauge symmetries of $U(1)_V \times U(1)_X$ are 
 broken,
 and remaining gauge symmetry becomes $SU(5)$. 
These VEVs of 4D Higgs fields 
 make unwanted fields become heavy through 
 the superpotential $W_{4D}^m$ in Eq.(\ref{25}). 
Then, only one set of ${\bf 10+\overline{5}+1}$ 
 is remaining in the low energy as the massless
 field. 
We should remind 
 that
 this chiral structure of the matter contents 
 is obtained because 
 ${\bf (27^f,27^f{}^c)}$ hyper-multiplet matter field 
 generates zero modes of ${\bf 10^f_{(-1,1)}}$ 
 and ${\bf \overline{5}^f_{(-2,2)}}$ in 
 Eqs.(\ref{19}) and (\ref{20}) 
 and ${\bf 1'_{x(0.0)}}$ in the 
 ${\bf (78',78'^c)}$ hyper-multiplet matter field 
 can not make mass term with 
 $\hat{\bf 45_0}$ in Eq.(\ref{25}). 
The four Higgs superfields in Eq.(\ref{14}) are also remaining 
 as the massless fields. 
When we denote the massless eigenstates as 
 ${\bf 10^0+\overline{5}^0+1^0}$, 
 they are written by the linear combinations of 
\beeq
{\bf 10^0}&\simeq&  
  \cos \phi_1 \cos \phi_3 \;{\bf 10_{(-1,-3)}} +
  \sin \phi_1 \cos \phi_3 \;{\bf 10_{(4,0)}} +
  \sin \phi_3 \;{\bf 10_{(-1,1)}^f} \nonumber \\ 
{\bf \overline{5}^0}&\simeq& 
   - \sin \theta_5 \;{\bf \overline{5}'{}_{(3,-3)}} 
   + \cos \theta_5 \;{\bf \overline{5}_{(-2,2)}^f},  \\
{\bf 1^0}&\simeq& 
   - \sin \theta_1 \;{\bf 1'{}_{(-5,-3)}} 
   + \cos \theta_1 \;{\bf 1'_{x(0,0)}} , \nonumber 
\eneq
with 
\beeq
&& \tan \phi_1 = 
  {\langle \hat{\bf 1}^h_{\bf 4}\rangle \over 
   \langle \hat{\bf 16}^h_{\bf 1}\rangle}
  {m_{{10}_2}m_{{10}^f_1}-m_{{10}_1}m_{{10}^f_2} \over 
   m_{{10}_1^c}m_{{10}^f_2}-m_{{10}_2^c}m_{{10}^f_1}}, \;\;\;\;\;
 \sin \phi_3 =
  {\langle \hat{\bf 1}^h_{\bf 4}\rangle 
   \langle \hat{\bf 16}^h_{\bf 1}\rangle
   (m_{{10}_1}m_{{10}^c_2}-m_{{10}_2}m_{{10}^c_1}) \over 
   \sqrt{A^2}} \nonumber \\
&&\tan \theta_5 = {m_{5'}\langle \hat{\bf 16}^h_{\bf 1}\rangle 
                   \over m_{5^f}}, \;\;\;\;\;\;\;\;\;\;
\;\;\;\;\;\;\;\;\;\;\;\;\;\;\;\;\;\;\;\;\;\;\;\;
  \tan \theta_1 = {m_{1'^c}
  \langle \hat{\bf 1}^h_{\bf 0}\rangle 
  \langle \hat{\bf 16}^h_{\bf 1}\rangle \over m_{1'}}, 
\eneq
where
$A^2 =
  \langle \hat{\bf 1}^h_{\bf 4} \rangle^2 
  \langle \hat{\bf 16}^h_{\bf 1}\rangle^2
   (m_{{10}_1}m_{{10}^c_2}-m_{{10}_2}m_{{10}^c_1})^2$ 
$+\langle \hat{\bf 1}^h_{\bf 4} \rangle^2
  (m_{{10}_2}m_{{10}^f_1}-m_{{10}_1}m_{{10}^f_2})^2$ 
$+\langle \hat{\bf 16}^h_{\bf 1}\rangle^2$ $
  (m_{{10}_1^c}m_{{10}^f_2}-m_{{10}_2^c}m_{{10}^f_1})^2$. 
Therefore 
 after breaking $U(1)_V \times U(1)_X$, 
 the Yukawa interactions of $SU(5)$, 
 $W_Y \equiv W_Y^a+W_Y^f$ in 
 Eqs.(\ref{y21}) and (\ref{27s}), become
\be
W^{eff}_Y = y_{u}{\bf 10^0\;10^0\;5^H} 
+ y_d {\bf 10^0\; \overline{5}^0\; \overline{5}^H}
+ y_\nu {\bf \overline{5}^0\; 1^0 \; 5^H}
+ y_M {\bf 1^H \; 1^0\; 1^0 },
\label{35}
\en
where 
\beeq
&& y_u \equiv {g \over2}\sin 2 \phi_1 \cos^2 \phi_3,\;\;\;\;\; 
   y_d \equiv -g \sin \phi_3 \sin \theta_5, \nonumber \\ 
&& y_\nu \equiv -g \sin \theta_5 \cos \theta_1, 
\;\;\;\;\;\;\;
   y_M \equiv -{g\over 2} \sin 2 \theta_1. 
\label{30n}
\eneq
When ${\bf 1^H}$ takes large VEV,  
 the 4th term of $W_Y^{eff}$ in Eq.(\ref{35}) induces the 
 Majorana mass of right-handed neutrinos.

Above discussions for generating the Yukawa interaction
 can be generalized to realistic three generation model
 by introducing a generation index to each bulk and 
 brane field. 
The introduction of 
 generation index $i$ $(i=1\sim3)$ 
 into Eqs.(\ref{y21}) and (\ref{27s}) 
 can induce the Yukawa interactions 
 of the three generations. 
As for the fermion mass hierarchies, 
 the suitable magnitudes can be obtained 
 by choosing the generalized angles in Eq.({\ref{30n}}). 
{}For examples, 
 the 2nd generation angles, 
 $(\sin 2 \phi_1)_2 \sim \lambda^4$ and 
 $(\sin \theta_5)_2 \sim \lambda^2$, and 
 the 1st generation angles, 
 $(\sin 2 \phi_1)_1 \sim \lambda^8$ and 
 $(\sin \theta_5)_1 \sim \lambda^6$ 
 can induce the suitable fermion 
 mass hierarchies\footnote{The suitable 
 mass squared differences of neutrinos can be also obtained
 by choosing the generalized angles in Eq.(\ref{30n}) 
 and Majorana mass scales.}, 
 where $\lambda$ stands for 
 the Cabibbo angle $\sim 0.2$. 
The order one angles of Eq.(\ref{30n}) 
 suggests that  
 the magnitudes of the Yukawa couplings 
 of the 3rd generation 
 are the same as those of the gauge couplings, which 
 is about $g \sim 0.7$ around the GUT scale. 
However, 
 we can not obtain the suitable 
 flavor mixings in the quark and lepton sectors 
 in this extension of introducing 
 the generation index $i$. 
It is because 
 the Yukawa interactions, 
 which are originated in the 5D gauge 
 interactions, 
 with generation index 
 in Eqs.(\ref{y21}) and (\ref{27s}) 
 are flavor diagonal. 
Thus, in order to obtain the suitable 
 flavor mixings of the quarks and leptons, 
 we should extend the brane fields and their 
 interactions 
 in Eq.(\ref{25}).
We add one more vector-like fields 
 $\hat{\bf 16}_{{\bf 1}} + 
  \overline{\hat{\bf 16}}_{{\bf -1}}$ for each generation 
 at $y=0$, 
 which means the index $k$ runs from 1 to 3 
 in Eq.(\ref{25}),  
 the superpotential at 4D wall 
 becomes 
\beeq
W_{4D}^m&=& 
     \sum_{k=1}^3\left(\overline{\hat{\bf 16}}_{{\bf -1}_k}\right)_i\;
    \;
     \left(\left(m_{{10}_k}\right)_{ij} \hat{\bf 1}_{\bf 4}^h \; 
     \left({\bf 16}_{{\bf -3}}\right)_j+ 
     \left( m_{{10^c}_k}\right)_{ij} \hat{\bf 16}^h_{\bf 1} 
      \left({\bf 45}_{{\bf 0}}\right)_j + 
      \left(m_{{10^f}_k}\right)_{ij}
      \left({\bf 16^f}_{\bf _1}\right)_j\right)
 \nonumber \\
&+&  \sum_{k=1}^3\left({\hat{\bf 16}}_{{\bf 1}_{k}}\right)_i\;
     \left(\left(\bar{m}_{{10}_k}\right)_{ij} \hat{\bf 1}_{\bf -4}^h \; 
              \left(\overline{\bf 16}_{{\bf 3}}\right)_j+ 
      \left(\bar{m}_{{10^c}_k}\right)_{ij} \overline{\hat{\bf 16}}^h_{\bf -1} 
             \left({\bf 45}_{{\bf 0}}\right)_j\right) \nonumber \\ % + 
&+& \left({\hat{\bf 10}}_{{\bf 2}}\right)_i\; 
     \left(\left(m_{5'}\right)_{ij} \hat{\bf 16}^h_{\bf 1}
     \left({\bf 16}_{{\bf -3}}\right)_j+ 
     \left( m_{5^f}\right)_{ij}
     \left({\bf  10^f}_{{\bf -2}}\right)_j\right)
\nonumber \\
&+&     \left(\hat{\bf 1}_{{\bf 0}}\right)_i\;
   \left(\left(m_{1'{}^c}\right)_{ij} \hat{\bf 1}_{\bf 4}^h \; 
   \overline{\hat{\bf 16}}^h_{\bf -1}
       \; \left({\bf 16'}_{{\bf -3}}\right)_j
  + \left(m_{1'}\right)_{ij} 
  \left({\bf 1'{}}_{{\bf 0}}\right)_j\right) \nonumber \\
&+& \left(\bar{m}_{5'} \right)_{ij}
  \left({\hat{\bf 10}}_{{\bf -2}}\right)_i 
  \overline{\hat{\bf 16}}^h_{\bf -1}
                \left( \overline{\bf 16}_{{\bf 3}}\right)_j+ 
    \left(\bar{m}_{1'{}^c}\right)_{ij} 
    \left(\hat{\bf 1}_{{\bf 0} }\right)_i
    {\hat{\bf 1}}_{\bf -4}^h \; 
                       {\hat{\bf 16}}^h_{\bf 1}
       \; \left(\overline{\bf 16}_{{\bf 3}}'\right)_j 
\\    
&+&\left(m_{24}\right)_{ij}\; \left(\hat{\bf 45}_{{\bf 0}}\right)_i\;
    \left({\bf 45}_{{\bf 0}}\right)_j
\label{W4d3} 
+    \sum_{k=1}^3 \left(M_{{10}_k}\right)_{ij} 
       \left(\overline{\hat{\bf 16}}_{{\bf -1}_k} \right)_i
             \left({\hat{\bf 16}}_{{\bf 1}_k} \right)_j
      + \left(M_{5}\right)_{ij} 
      \left({\hat{\bf 10}}_{{\bf 2}}\right)_i
      \left({\hat{\bf 10}}_{{\bf -2}}\right)_j
\nonumber\\
      &+& \left(M_{1}\right)_{ij} 
      \left(\hat{\bf 1}_{{\bf 0}}\right)_i
      \left(\hat{\bf 1}_{{\bf 0}}\right)_j
      +\left( M_{24}\right)_{ij}\; 
      \left(\hat{\bf 45}_{{\bf 0}}\right)_i
      \left(\hat{\bf 45}_{{\bf 0}}\right)_j,
\nonumber 
\eneq
where $i,j$ stand for the generation indexes. 
Contrary to the case of Eq.(\ref{25}), 
 massless {\bf 10} representation fields 
 appear mainly from the combinations of 
 brane fields, $\hat{\bf 10}$s, 
 when $(M_{10})_{ij} \ll (m_{10})_{ij}, (m_{{10}^c})_{ij}, 
 (m_{{10}^f})_{ij},(\bar{m}_{10})_{ij}$.  
It can be easily shown 
 by considering 
 the vanishing limit of $(M_{10})_{ij}$, 
 where 
 massless fields appear only from 
 the combinations of $\hat{\bf 10}$s.

Now let us 
 assume that $(M_{10})_{ij}=(M_{10})_i\delta_{ij}$, 
 and 
 the 4D Higgs fields can take VEVs 
 of order $\Lambda$ 
 through the superpotential,  
\beeq
W_{4D}^h= Y_1(\hat{\bf 16}^h_{\bf 1} 
  \overline{\hat{\bf 16}}^h_{\bf -1}-\Lambda^2 )+
  Y_2(\hat{\bf 1}^h_{\bf -4} \hat{1}^h_{\bf 4}-\Lambda^2 ),
\eneq
at $y=0$ wall, where
 $Y_l$s $(l=1,2)$ are gauge singlet fields. 
We also assume that 
 $O(\Lambda (m_{{10}_k})_{ij})
 =O(\Lambda (m_{{10^c}_k})_{ij})
 =O((m_{{10}^f_k})_{ij})
 =O(\Lambda (\overline{m}_{{10}_k})_{ij})
 =O(\Lambda (\overline{m}_{{10^c}_k})_{ij})\equiv m$.
Then, after the breaking of 
 $U(1)_V \times U(1)_X$, 
 a massless {\bf 10} representation field of 
 the $i$-th generation 
 can be written by 
\beeq
\left({\bf 10}^{\bf 0}\right)_{i}  &\sim &  
\frac{\left(M_{10}\right)_i}{m}\left(
\left(c_{1}\right)_{ij} \left({\bf 10}_{{\bf (-1,-3)}}\right)_{j}+
\left(c_{2}\right)_{ij} \left({\bf 10}_{{\bf (4,0)}}\right)_j
+\left(c_{3}\right)_{ij} \left({\bf 10^f}_{{\bf(-1,1)}}\right)_j
\right)
\nonumber \\
&+& 
\left(c_{4}\right)_{ij} \left({\bf \hat{10}}_{{\bf (4,0)}}\right)_j
+\left(c_{5}\right)_{ij} \left({\bf \hat{10}}_{{(4,0)}}\right)_j
+ \left(c_{6}\right)_{ij}\left({\bf \hat{10}}_{{(4,0)}}\right)_j
\label{36},
\eneq
when $m \gg (M_{10})_i$,
where $(c_{ij})$s are O(1) coefficients with no hierarchies. 
By taking the parameters, 
 $(M_{10})_1/{m} \sim \lambda^4$, 
 $(M_{10})_2/{m} \sim \lambda^2$, and 
 $(M_{10})_3/{m} \sim 1$, 
 the bulk fields can have the zero mode components as 
\beeq
{\bf 10}_{{\bf (-1,-3)} } \simeq V {\bf 10}^{\bf 0}_{ },~~
{\bf 10}_{{\bf (4,0)}} \simeq V  {\bf 10}^{\bf 0}_{ },~~
{\bf 10^f}_{{\bf (-1,1)}} \simeq V {\bf 10}^{\bf 0}_{},
\label{zeromode}
\eneq
with 
\begin{equation}
 V \simeq \left(
\begin{array}{ccc}
 \lambda^4 & \lambda^4 &  \lambda^4  \\
 \lambda^2 & \lambda^2  &  \lambda^2  \\
 1  & 1  & 1
\end{array}
\right)
\end{equation}
in the matrix form 
where we suppress the $O(1)$ coefficients for each component.
Then, the Yukawa interactions in Eqs.(19) and 
 (24) with 
 the generation index and 
 Eqs.(31) and (\ref{zeromode})
 suggest the 
 Yukawa interactions 
\be
W^{eff}_Y = y_{u}{\bf 10^0\;10^0\;5^H} 
+ y_d {\bf 10^0\; \overline{5}^0\; \overline{5}^H}
+ y_\nu {\bf \overline{5}^0\; 1^0 \; 5^H}
+ y_M {\bf 1^H \; 1^0\; 1^0 }
%\label{35}
\en
in the low energy
 with the Yukawa couplings, 
\begin{equation}
 y_u^{} \simeq \left(
\begin{array}{ccc}
 \lambda^8 & \lambda^6 &  \lambda^4  \\
 \lambda^6 & \lambda^4  &  \lambda^2  \\
 \lambda^4  & \lambda^2  & 1
\end{array}
\right)  , \;\;
 y_d^{} \simeq \left(
\begin{array}{ccc}
 \lambda^4 & \lambda^4 &  \lambda^4  \\
 \lambda^2 & \lambda^2  &  \lambda^2  \\
  1 & 1 & 1 
\end{array}
\right) , \;\;
 y_\nu^{},  y_M^{} \simeq \left(
\begin{array}{ccc}
1 & 1& 1\\
1 & 1& 1\\
1 & 1& 1
\end{array}
\right). 
\label{mass}
\end{equation}
These mass matrices  
 can be suitable for the phenomenology\cite{BB,Haba:2000be}. 
They can induce the small (large) flavor mixings in the 
 quark (lepton) sector as well as 
 the suitable 
 fermion mass hierarchies.  
Moreover, the CP violating phase can arise from 
 the $O(1)$ complex coefficients
 of the Yukawa couplings in
 Eq.(\ref{mass}).  
It is because 
 the brane-localized interactions in Eq.(31) 
 have the complex couplings, 
 where the physical CP phase can not be 
 rotated out. 
Thus, we can obtain the physical CP violating phase 
 in the Yukawa sector, although all Yukawa 
 interactions 
 come from the 5D gauge interactions with 
 no CP phases in origin.

In the mass matrices of Eq.(37), 
 the suitable choices of $O(1)$ coefficients 
 are needed for the 
 suitable magnitudes of $V_{us}$, $U_{e3}$,  
 and down quark and electron 
 masses\footnote{ 
If $O(1)$ coefficients are not determined by 
 a specific reason (symmetry) in the fundamental theory, 
 it is meaningful to 
 see the most probable hierarchies and mixing angles 
 by considering random $O(1)$ coefficients \cite{Haba:2000be}.}. 
So if the fermion mass hierarchies and 
 flavor mixing angles should determined from the fundamental 
 theory 
 in {\it order} (power of $\lambda$) not by tunings of 
 $O(1)$ coefficients, 
 we should modify this scenario. 
But, we can make this modification in a very simple way. 
We introduce one more 
 vector-like $\hat{\bf 10}_{\bf 2}+{\hat{\bf 10}}_{\bf -2}$ 
 fields for each generation at $y=0$. 
Then, the parameters in Eq.(31) should be 
 changed as 
 $(m_{5'})_{ij}\rightarrow (m_{{5'}_l})_{ij}$, 
 $(m_{5^f})_{ij}\rightarrow (m_{{5^f}_l})_{ij}$, 
 $(\bar{m}_{5'})_{ij}\rightarrow (\bar{m}_{{5'}_l})_{ij}$, 
 and  
 $(M_5)_{ij}\rightarrow (M_{{5}_l})_{ij}$
 with $l=1,2$. 
Assuming 
 $O((m_{{5^f}_l})_{ij})=
 O(\Lambda (m_{{5'}_l})_{ij})=
 O((\bar{m}_{{5'}_l})_{ij})= m_5$, 
 $(M_{{5}})_{ij}=(M_{5})_i\delta_{ij}$, and 
 $(M_5)_1/(m_5)_1\sim \lambda^2$,  
 $(M_5)_{2,3}/(m_5)_{2,3}\sim 1$, 
 the zero mode field of the 1st generation 
 ${\bf \overline{5}}$ field 
 accompanies with $\lambda^2$ as,
 $\lambda^2 ({\bf \overline{5}}^{\bf 0})_{1}$, 
 similar to the case of 
 {\bf 10} representations. 
By this modification, we can obtain 
 the modified mass matrices of quarks and leptons, 
 which can induce the suitable values of 
 $V_{us}$, $U_{e3}$, and 
 down quark and electron masses
 (this type of mass matrices is the 
 modification I in Ref.\cite{HKS2}). 
In the same way, 
 we can obtain 
 mass matrices of the small $\tan \beta$ case 
 (modification II in Ref.\cite{HKS2})
 by choosing the parameters of 
 $(M_{10})_i/m$ and $(M_5)_i/m_5$.

Our mechanism of generating 
 the fermion mass hierarchies and flavor
 mixings is very simple, where we 
 can also obtain the CP violation 
 in the effective 4D Yukawa couplings. 
In this point our mechanism 
 is different from that of 
 Ref.\cite{Burdman:2002se} in which 
 generating 
 Yukawa hierarchies is discussed by 
 the wave function 
 localization.

%%%%%%%%%%%%%%%%%%%%%%%%%%%%%%%%%%%%%%%%%%%%%%%%%%%%%%%%%%%%%%%%%%%
%%%%%%%%%%%%%            Section            %%%%%%%%%%%%%%%%%%%%%%%
%%%%%%%%%%%%%%%%%%%%%%%%%%%%%%%%%%%%%%%%%%%%%%%%%%%%%%%%%%%%%%%%%%%
\section{$E_7$ GUT}

Next we will consider $E_7$ GUT. 
The $E_6$ GUT discussed in the previous section
 must have two different representations, 
 the fundamental and adjoint representations,
where the opposite parity eigenstate of 
 the adjoint representation is also needed 
 in order to generate all Yukawa interactions of 
 quarks and leptons. 
However, in $E_7$ GUT, 
 this situation is improved, where 
 we can obtain all Yukawa interactions only from an
 adjoint representation matter hyper-multiplet in the bulk.

The adjoint representation of $E_7$ is 
 {\bf 133}, which is divided as,
\beeq
&&{\bf 133 = (66,1)+(1,3)+(32',2)},
       \;\;\;\;\;\;\;\;\;\;\;\;\;\;\;\;\;\;SO(12)\times SU(2), 
\label{40}  \\
&&{\bf 133 = 78_0 + 1_0 + 27_{1}+\overline{27}_{-1}},
     \;\;\;\;\;\;\;\;\;\;\;\;\;\;\;\;\;\;
     \;\;\;E_6\times U(1)_Z .
\label{41} 
\eneq
We take parity $P$ and boundary conditions $T$ 
 as the reductions of $E_7$ 
 into $E_6 \times U(1)_Z$ and 
 $SO(12)\times SU(2)$,
 respectively.
As for vector hyper-multiplets, 
 we take $P=+1$ for ${\bf 78_0+1_0}$, and
 $P=-1$ for ${\bf 27_1+\overline{27}_{-1}}$ in Eq.(39). 
On the other hand, 
 we take $T=+1$ for ${\bf (66,1)+(1,3)}$, 
 and $T=-1$ for 
 ${\bf (32',2)}$ 
 in Eq.(38). 
These $P$ and $T$ make 
 the $E_7$ gauge symmetry reduce to 
 $SO(10) \times U(1)_X \times U(1)_Z$. 
It is because 
 the vector superfield, $V_{\bf 133}$, is divided into 
\beeq
V_{\bf 133}&=&
  {\bf 45_{(0,0)}^{(+,+)} + 1_{x(0,0)}^{(+,+)} + 1_{z(0,0)}^{(+,+)} 
     + 16_{(-3,0)}^{(+,-)}+\overline{16}_{(3,0)}^{(+,-)}
     + 10_{(-2,1)}^{(-,+)}+\overline{10}_{(2,-1)}^{(-,+)}} \nonumber \\
&& +{\bf 1_{(4,1)}^{(-,+)}+1_{(-4,-1)}^{(-,+)}
     + 16_{(1,1)}^{(-,-)}+\overline{16}_{(-1,-1)}^{(-,-)}}
\label{43}
\eneq
under $SO(10) \times U(1)_X \times U(1)_Z$. 
Here $(\pm, \pm)$ shows the eigenvalues of 
 $(P,T)$, and the charges show 
 the quantum numbers of 
 $SO(10) \times U(1)_X \times U(1)_Z$.  
The zero modes, $(+,+)$ elements, in $V_{\bf 133}$ show that 
 the gauge symmetry is reduced as 
 $E_7 \rightarrow SO(10) \times U(1)_X \times U(1)_Z$.  
While 
 the chiral superfield 
 $\Sigma_{\bf 133}$ is divided into 
\beeq
\Sigma_{\bf 133}&=&
  {\bf 45_{(0,0)}^{(-,+)} + 1_{x(0,0)}^{(-,+)} + 1_{z(0,0)}^{(-,+)} 
     + 16_{(-3,0)}^{(-,-)}+\overline{16}_{(3,0)}^{(-,-)}
     + 10_{(-2,1)}^{(+,+)}+\overline{10}_{(2,-1)}^{(+,+)}} \nonumber \\
&& +{\bf 1_{(4,1)}^{(+,+)}+1_{(-4,-1)}^{(+,+)}
     + 16_{(1,1)}^{(+,-)}+\overline{16}_{(-1,-1)}^{(+,-)}}.
\label{44}
\eneq
The zero modes, $(+,+)$, suggest that
 there appear the following four Higgs fields, 
\be
{\bf 10_{(-2,1)}^{H}, \; 10_{(2,-1)}^{H}, \;
       1_{(4,1)}^{H}, \; 1_{(-4,-1)}^{H}},
\label{sigma}
\en
in the low energy, 
 since we regard $\Sigma$ as the Higgs fields
 as in Eq.(\ref{Yukawa}).

\begin{figure}
\begin{center}
\epsfig{file=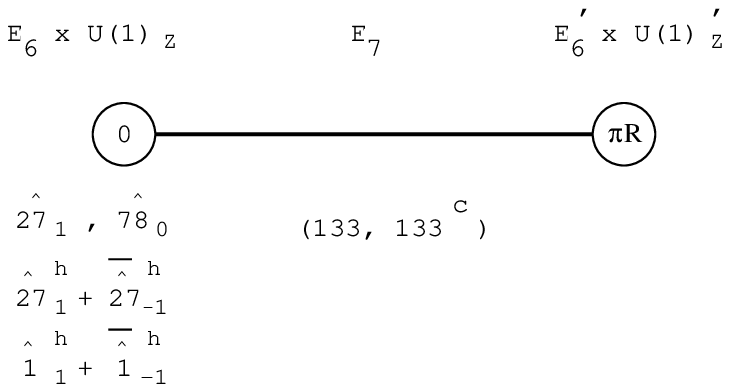,width=10cm}
\caption{Matter configuration on the $S^1/Z_2$ orbifold in the $E_7$
 GUT. }
\label{fig:E7}
\end{center}
\end{figure}

As for the matter fields, we consider the
 adjoint representation 
 hyper-multiplet, $({\bf 133}, {\bf 133^c})$.  
In Fig.\ref{fig:E7} we have shown the matter configuration
 on the orbifold. 
We choose the eigenstates of $P$ and $T$ as, 
\beeq
{\bf 133}&=&
  {\bf 45_{(0,0)}^{(+,-)} + 1_{x(0,0)}^{(+,-)} + 1_{z(0,0)}^{(+,-)} 
     + 16_{(-3,0)}^{(+,+)}+\overline{16}_{(3,0)}^{(+,+)}
     + 10_{(-2,1)}^{(-,-)}+\overline{10}_{(2,-1)}^{(-,-)}} \nonumber \\
&& +{\bf 1_{(4,1)}^{(-,-)}+1_{(-4,-1)}^{(-,-)}
     + 16_{(1,1)}^{(-,+)}+\overline{16}_{(-1,-1)}^{(-,+)}}
\label{45} \\
{\bf 133^c}&=&
  {\bf 45_{(0,0)}^{(-,-)} + 1_{x(0,0)}^{(-,-)} + 1_{z(0,0)}^{(-,-)} 
     + 16_{(-3,0)}^{(-,+)}+\overline{16}_{(3,0)}^{(-,+)}
     + 10_{(-2,1)}^{(+,-)}+\overline{10}_{(2,-1)}^{(+,-)}} \nonumber \\
&& +{\bf 1_{(4,1)}^{(+,-)}+1_{(-4,-1)}^{(+,-)}
     + 16_{(1,1)}^{(+,+)}+\overline{16}_{(-1,-1)}^{(+,+)}} .
\label{46}
\eneq
Then the Yukawa interactions in the low energy Eq.(\ref{Yukawa}) 
 become
\beeq
{\bf 133^c}\; \Sigma_{\bf 133}\; {\bf 133}\; & \sim & 
\;{\bf 16_{(1,1)}\; 10_{(2,-1)}^H \; 16_{(-3,0)}+
      \overline{16}_{(-1,-1)}\; 10_{(-2,1)}^H \; 
                               \overline{16}_{(3,0)}} \nonumber \\ 
&&  +{\bf  16_{(1,1)}\; 1_{(-4,-1)}^H \; \overline{16}_{(3,0)} 
    + \overline{16}_{(-1,-1)}\; 1_{(4,1)}^H \; 16_{(-3,0)}} 
\label{47}
\eneq
{}from Eqs.(41), (\ref{45}), and (\ref{46}). 
This case can not have the chiral 
 matter structure, since 
 we only introduce the adjoint matter field. 
Of cause we can obtain the chiral 
 matter structure as the 
 $E_6$ GUT case if 
 we introduce the fundamental representation 
 matter fields.
But here we would suggest another possibility
 of the scenario. 
That is, introducing 
 the 4D brane field, $\hat{\bf 27_1}$, at 
 $y=0$ wall. 
Then the field ${\bf \overline{16}_{(-1,-1)}}$ 
 in Eq.(45) becomes heavy 
 through the 
 brane-localized mixing mass 
 with 
 $\hat{\bf 27_1}\supset \hat{\bf 16}_{\bf (1,1)}$, 
 and can not survive 
 in the low energy. 
So the 2nd and 4th terms in Eq(45) become irrelevant. 
However, at the 4D wall $y=0$, 
 there exist gauge anomalies related with 
 $U(1)_Z$. 
We assume 
 these anomalies can be canceled 
 out by the Green-Schwarz mechanism\cite{GS} 
 on the 4D wall\footnote{If we take this 
 assumption in the $E_6$ GUT, 
 the matter contents in the previous 
 section can be simple. }, which can be 
 available since  
 $E_6$ is anomaly free.

As for the breaking of $U(1)_X \times U(1)_Z$, 
 we introduce Higgs fields which transform 
 $\hat{\bf 27}^h_{\bf 1} + \overline{\hat{\bf 27}}^h_{\bf -1}$ and 
 $\hat{\bf 1}^h_{\bf 1} + {\hat{\bf 1}}^h_{\bf -1}$ 
 under $E_6 \times U(1)_Z$ on the 4D wall at $y=0$. 
We assume that they take VEVs 
 around the GUT scale as, 
 $\langle \hat{\bf 27}^h_{\bf 1}\rangle \simeq 
  \langle \hat{\bf 1}^h_{\bf {(4,1)}}\rangle$, 
 $\langle \overline{\hat{\bf 27}}^h_{\bf -1}\rangle \simeq 
  \langle \hat{\bf 1}^h_{\bf {(-4,-1)}}\rangle$,
 $\langle \hat{\bf 1}^h_{\bf 1}\rangle \simeq 
  \langle \hat{\bf 1}^h_{\bf (0,1)}\rangle$, and 
 $\langle \hat{\bf 1}^h_{\bf -1}\rangle \simeq
  \langle \hat{\bf 1}^h_{\bf (0,-1)}\rangle$. 
We also introduce 4D brane-localized matter fields, 
 $\hat{\bf 78}_{\bf 0}$, 
 at $y=0$. 
Then the superpotential $W^m_{4D}$ is given by 
\beeq
W^m_{4D}&=& \hat{\bf 78}_{\bf 0}
     (m_E \overline{\hat{\bf 27}}^h_{\bf -1}{\bf 27_1}  
    + m_E'{\bf 78_0}) +M_E \hat{\bf 78}_{\bf 0}^2,
%\label{48}
\eneq
which means that 
 the massless field is given by 
\beeq
{\bf 16^0}&\simeq&  
   -\sin \theta_{16} {\bf 16_{(1,1)}} 
   +\cos \theta_{16} {\bf 16_{(-3,0)}},
\eneq
with 
\beeq
\tan \theta_{16} = {m_E' /  
   (m_E \langle  \overline{\hat{\bf 27}}^h_{\bf -1}\rangle)}.
\eneq
As the consequence, we can obtain the 
 Yukawa interaction of $SO(10)$ from Eq.(\ref{47}) as,  
\be
W_Y^{eff}\simeq  y_{10}{\bf 16^0\;16^0\;10^H},
\label{50}
\en
with $y_{10}\equiv -2 \sin 2 \theta_{16}$ 
 below the energy scale of $U(1)_X \times U(1)_Z$ 
 breaking. 
It is worth noting that all Yukawa interactions 
 of SUSY SM 
 can be generated by only a single representation of 
 matter hyper-multiplet. 
This is the significant feature of $E_7$ GUT.  
However, for extending three-generation realistic model, 
 we can not use the same technique of generating 
 the suitable 
 fermion mass hierarchies and flavor mixings 
 as in
 the $E_6$ case, 
 since the $SO(10)$ gauge symmetry is 
 remaining and {\bf 10} representation of $SU(5)$
 can not be treated separately. 
Besides, 
 we can not obtain the CKM matrix from 
 Eq.(\ref{50}) because of the absence of 
 another Yukawa coupling, 
 $y_{10}'{\bf 16\;16\;10'^H}$, in $W_Y^{eff}$. 
Thus, we should consider other 4D mechanisms 
 of generating fermion mass hierarchies and flavor mixings 
 below the 
 $SO(10)$ gauge breaking\footnote{The $SO(10)$ can be broken 
 by introducing additional 4D brane-localized 
 Higgs field, $\hat{\bf 78}_{\bf 0}^h$ at $y=0$.}, 
 which is nothing to do with the orbifolding.

Before closing this section, 
 we comment on the cases of other choices 
 of $P$ and $T$ rather than Eqs.(\ref{45}) and (\ref{46}). 
{}For realizing the suitable gauge reduction,
 $E_7 \rightarrow SO(10) \times U(1)_X \times U(1)_Z$, 
 there are two other choices of 
 $P$ and $T$. 
One is taking $P=+1$ for 
 ${\bf (66,1)+(1,3)}$ and $P=-1$ 
 for ${\bf (32',2)}$ in Eq.(\ref{40}), and 
 $T=+1$ for ${\bf 78_0+1_0}$, and 
 $T=-1$ for ${\bf 27_1+\overline{27}_{-1}}$ in Eq.(\ref{41}). 
In this case $V$ has zero modes in the components of 
 ${\bf 45_{(0,0)}+1_{x(0,0)}+1_{z(0,0)}}$, while 
 $\Sigma$, which becomes Higgs fields, 
 has zero modes at 
 ${\bf 16_{(-3,0)}+\overline{16}_{(3,0)}}$ 
 in Eq.(41). 
This case  
 can not produce
 Yukawa interactions of down-sector and 
 charged lepton by adjoint representation
 matter fields in the bulk. 
{}For the down-quark and charged lepton Yukawa
 interactions, we should introduce
 fundamental representation matters in the bulk.  
The other is taking $P=+1$ for 
 ${\bf 78_0 + 1_0}$ and 
 $P=-1$ for ${\bf 27_{1}+\overline{27}_{-1}}$, 
 and $P'=+1$ for ${\bf (66,1)+(1,3)}$ and
 $P'=-1$ for ${\bf (32',2)}$, 
 where $P' \equiv TP$. 
This case suggests 
 the residual gauge symmetry at 
 $y=0$ is $E_6\times U(1)_Z$ 
 and that at $y=\pi R$ is $SO(12)\times SU(2)$.   
The Higgs fields are 
 ${\bf 16_{(1,1)}+\overline{16}_{(-1,-1)}}$  
 in this case. 
This case can not also induce 
 Yukawa interactions of down-sector quarks and 
 charged leptons in 
 Eq.(\ref{Yukawa}) by  
 adjoint representation fields. 
Thus, 
 no other choices of $P$ and $T$ 
 except for Eqs.(\ref{45}) and (\ref{46}) can induce 
 all Yukawa interactions of 
 quarks and leptons by only 
 single matter representation\footnote{We
 do not consider matter representations larger than 
 the adjoint representation. 
 Such higher representations 
 produce a lot of unwanted fields in the low 
 energy, and also induce sudden blow-up of the 
 gauge couplings.}. 
This situation is not changed if 
 we take $E_7 \supset SU(6) \times SU(3)$ for 
 $P$ and $T$, where ${\bf 133 = (35,1)+(1,8)+
 (15,\overline{3})+(\overline{15},3)}$. 
This case also needs both the adjoint and fundamental
 representations for generating all Yukawa interactions 
 of quarks and leptons in the low energy.

%%%%%%%%%%%%%%%%%%%%%%%%%%%%%%%%%%%%%%%%%%%%%%%%%%%%%%%%%%%%%%%%%
%%%%%%%%%%%%  Summary     %%%%%%%%%%%%%%%%%%%%%%%%%%%%%%%%%%%%%%%
%%%%%%%%%%%%%%%%%%%%%%%%%%%%%%%%%%%%%%%%%%%%%%%%%%%%%%%%%%%%%%%%% 
\section{Summary and Discussion}

We have considered 5D 
 $E_6$ and $E_7$ 
 supersymmetric GUTs on 
 the orbifold. 
We adopt such large gauge groups 
 in order to respect 
 the charge quantization and 
 the simple anomaly free structure on the 4D fixed wall. 
A small numbers of matter representations 
 are enough for inducing all Yukawa 
 interactions of the SM. 
The gauge groups are reduced to their subgroups by imposing 
 the non-trivial parity and boundary conditions for the gauge fields.
Moreover, the suitable Higgs fields can arise from the 
 5th components of the 5D gauge fields. 
We have introduced bulk matter fields so that 
 the Yukawa interactions are induced from 
 the gauge interactions 
 with the 5th component of gauge fields, 
 which are identified with the Higgs fields. 
In the $E_6$ GUT, 
 we have introduced the adjoint and 
 fundamental representation hyper-multiplets, 
 ${\bf (78^c,78)}$ and ${\bf (27^c,27)}$, respectively, 
 for the bulk matter fields. 
The Yukawa interactions for up-sector quarks and Dirac- and
 Majorana-neutrino masses are induced from 
 the gauge interactions of
 the ${\bf (78^c,78)}$ matter hyper-multiplets, 
 and those for down-sector quarks 
 and charged leptons
 are induced from the gauge interactions of 
 ${\bf (27^c,27)}$ matter hyper-multiplet.
For the phenomenologically acceptable model, 
 we have introduced additional 
 matters and Higgs fields on the 4D brane 
 with anomaly free contents. 
The vacuum expectation values of the 
 brane-localized Higgs fields 
 have made unwanted matter fields heavy as well as 
 generate effective Yukawa interactions. 
Thanks for the brane-localized mass terms 
 between the bulk and brane-localized matter 
 fields, 
 the realistic fermion 
 mass hierarchies and flavor mixings 
 can be produced 
 by integrating out heavy fields. 
As for the 
 $E_7$ GUT, 
 all Yukawa interactions can be produced 
 by only single representation of 
 matter hyper-multiplets, ${\bf (133^c,133)}$ and 
 ${\bf (248^c,248)}$, respectively.

Here
 let us discuss the origins of the $\mu$ term in our scenario.
As discussed in Ref.\cite{Burdman:2002se}, 
 we can consider two mechanisms for 
 generating the $\mu$ term. 
One is the supergravity effects\cite{Giudice:1988yz}, 
 where the term $[\Sigma^2]_D$, which is induced 
 in 4D K\"ahler potential from the 5D gauge kinetic term, 
 can lead to $\mu$ term. 
The existence of $U(1)_R$ symmetry in the 5 dimension  
 forbids the direct $\mu$ term as well as
 the dangerous dimension 4 and 5 proton decay operators, 
 where the Higgs fields have no $R$ charges. 
The other is orbifolding 
 boundary effect\cite{SS}, where  
 the gaugino masses are induced through 
 the SUSY breaking boundary conditions. 
The $\mu$ term can 
 be generated in the same mechanism since the Higgs fields
 arise from the gauge fields in higher dimensions.

{}Finally let us comment on the triplet-doublet (TD) splitting 
 in our models. 
When we consider the Wilson line operator, 
 ${\mathcal P}\exp( \int \Sigma dy)$, we can 
 consider the brane-localized gauge 
 invariant interactions\cite{Hall:2001zb,Burdman:2002se},
 where we can adopt the usual 
 4D solutions of TD splitting 
 problem\cite{missing,pG,DW}. 
Here ${\mathcal P}$ shows the path ordered product. 
On the other hand, 
 when we permit the higher representation 
 matter fields, the missing partner 
 mechanism\cite{missing}  
 can work in our scenarios. 
{}For examples, 
 in $E_6$ GUT, 
 we introduce Higgs hyper-multiplets, 
 $({\bf 1728},{\bf 1728}^c)$ and 
 $(\overline{\bf 1728},\overline{\bf 1728}^c)$ 
 in the bulk. 
The orbifolding in the section 
 3 shows 
 ${\bf 1728}^c \supset {\bf 75_{(0,-4)}^{(+,+)}}$, 
 ${\bf 1728} \supset \overline{\bf 50}_{\bf (3,1)}^{(+,+)}$, 
 $\overline{\bf 1728}^c \supset {\bf 75_{(0,4)}^{(+,+)}}$, and
 $\overline{\bf 1728} \supset {\bf 50}_{\bf (-3,1)}^{(+,+)}$. 
Thus, the gauge interactions in Eq.(\ref{Yukawa}), 
 ${\bf 1728}^c$ $\Sigma_{\bf 78}$ ${\bf 1728}$ and 
 $\overline{\bf 1728}^c$ $\Sigma_{\bf 78}$ 
 $\overline{\bf 1728}$, 
 induce the Yukawa interactions, 
 ${\bf 75_{(0,-4)}}\;{\bf 5^H_{(-3,3)}}\;
  {\bf \overline{50}}_{\bf (3,1)}$ and
 ${\bf 75_{(0,4)}}\;{\bf \overline{5}^H}_{\bf (3,-3)}\;
  {\bf 50^c_{(-3,1)}}$, respectively. 
Below the breaking of $U(1)_V \times U(1)_X$, 
 only the triplet can take GUT scale mass, 
 through the GUT scale VEV of ${\bf (1,1)_0}$ component 
 of {\bf 75}s under $(SU(3)_c,SU(2)_L)_{U(1)_Y}$. 
It is because 
 ${\bf 50\, (\overline{50})}$ contains 
 ${\bf (3,1)_{-2} \,((\overline{3},1)_2)}$ 
 but not contains ${\bf (1,2)_3 \,((1,2)_{-3})}$. 
This is just the usual missing partner 
 mechanism\cite{missing}.

%%%%%%%%%%%%%%%%%%%%%%%%%%%%%%%%%%%%%%%%%%%%%%%%%%%%%%%%%%%%%%%%%%%
%%%%%%%%%%%%%       Acknowledgment          %%%%%%%%%%%%%%%%%%%%%%%
%%%%%%%%%%%%%%%%%%%%%%%%%%%%%%%%%%%%%%%%%%%%%%%%%%%%%%%%%%%%%%%%%%%

\section*{Acknowledgment}
We would like to thank Y. Kawamura, Y. Hosotani, 
 K. Inoue, Y. Nomura, N. Maru, N. Yamashita, and T. Ota for 
 helpful discussions. 
This work is supported in
 part by the Grant-in-Aid for Science
 Research, Ministry of Education, Culture, Sports, Science and 
 Technology, of Japan (No. 14039207, No. 14046208, No. 14740164).

%%%%%%%%%%%%%%%%%%%%%%%%%%%%%%%%%%%%%%%%%%%%%%%%%%%%%%%%%%%%%%%%%
%%%%%%%%%%%%  References  %%%%%%%%%%%%%%%%%%%%%%%%%%%%%%%%%%%%%%%
%%%%%%%%%%%%%%%%%%%%%%%%%%%%%%%%%%%%%%%%%%%%%%%%%%%%%%%%%%%%%%%%%

\vskip 1.5cm

\leftline{\bf Note added}

The first version of this paper 
 contained the $E_8$ GUT, where 
 $E_6$ GUT is said to be realized by two 
 $Z_2$ transformations. 
However, 
 we become aware of the 
 paper that studies orbifolding 
 in $E_8$\cite{note1}, which shows  
 $E_8 \rightarrow E_6 \times SU(3)$ 
 can not be realized by $Z_2$ parity. 
It is because the assignment of 
 ${\bf (27,3)}$ and 
 ${\bf (\overline{27}, \overline{3})}$ 
 to have a negative parity is not 
 consistent, since 
 $[{\bf (27,3)}, {\bf (27,3)}] \subset 
   {\bf (\overline{27}, \overline{3})}$.  
% which is not correspond to an algebra 
% automorphism. 
To realize the breaking of 
 $E_8 \rightarrow E_6 \times SU(3)$, 
 $Z_3$ orbifolding is needed, for examples, 
 in ``two'' extra dimensional theory\cite{note2}.

\end{document}